\documentclass{ptephy_v1}
\preprintnumber{OCU-PHYS-505, AP-GR-155, WU-AP/1905/19} 
\usepackage[utf8]{inputenc}
\usepackage{latexsym,array,theorem,mathrsfs,bm,float}
\usepackage{psfrag}
\usepackage{amsfonts,amsmath,amssymb,latexsym,array,afterpage,theorem,color,graphicx,mathrsfs,bm,float,tabularx,here}
\usepackage{multirow}
\newcommand{\bea}{\begin{eqnarray}}
\newcommand{\ena}{\end{eqnarray}}
\newcommand{\beann}{\begin{eqnarray*}}
\newcommand{\enann}{\end{eqnarray*}}

\newcommand{\f}[2]{\frac{#1}{#2}}
\begin{document}
\title{
Maximal Efficiency of Collisional Penrose Process with Spinning Particle II 
\\
\normalsize{Collision with a Particle on ISCO}
}


\author{Kazumasa Okabayashi}
\affil{Department of Mathematics and Physics, Osaka City University, Osaka 558-8585, Japan 
\email{okabayashi"at"ka.osaka-cu.ac.jp}
}

\author[2]{Kei-ichi Maeda}
\affil{Department of Physics, Waseda University,
Shinjuku, Tokyo 169-8555, Japan}
\affil{Waseda Institute for Advanced Study {\rm (WIAS)}, 1-6-1 Nishi Waseda, Shinjuku-ku, Tokyo 169-8050, JAPAN
\email{maeda"at"waseda.jp}
}


\begin{abstract}
We analyze the collisional Penrose process between a particle on the ISCO orbit around an extreme Kerr black hole and a particle impinging from infinity.
We consider both cases with non-spinning and spinning particles.
We evaluate the maximal efficiency, $\eta_{\text{max}}=(\text{extracted energy})/(\text{input energy})$, for the elastic collision of two massive particles and for the photoemission process,
in which the ISCO particle will escape to infinity after a collision with a massless impinging particle. 
For non-spinning particles, the maximum efficiency is 
$\eta_{\text{max}} \approx 2.562$ for the elastic collision and 
 $\eta_{\text{max}} \approx 7$ 
 for the photoemission process.
For spinning particles we obtain  the maximal efficiency
 $\eta_{\text{max}} \approx 8.442$ for the elastic collision and 
 $\eta_{\text{max}} \approx 12.54$  for the photoemission process.
\end{abstract}

\subjectindex{}

\maketitle

\section{INTRODUCTION}
Energy extraction from a rotating black hole is 
one of the most fundamental issues in black hole physics.
It may also be important for relativistic astrophysics.
The Penrose process, in which we can extract the rotational energy using an ergo region of 
a black hole plays a key role in the extraction.
In particular, the collisional Penrose process may be more important because 
it may give more efficient energy extraction.
Although it has been pointed out that the center of mass energy diverges in the collision at the horizon of an extreme Kerr back hole between two particles impinging from infinity (BSW effect) \cite{Banados:2009pr},
it does not mean that one can exact as much energy as one would like 
because the infinite energy near the horizon will be red-shifted away. 
Hence, it is important to calculate how much energy we can really extract to infinity.
The  maximal  efficiency $\eta=(\text{extracted energy})/(\text{input energy})$
has been calculated in collisions between particles impinging from infinity by many authors\cite{Bejger:2012yb,Harada:2012ap,Schnittman:2014zsa,
Leiderschneider:2015kwa,Ogasawara:2015umo,Zaslavskii:2016unn,Mukherjee2019}. 
As for BSW effect, there have also been so far many studies\cite{Jacobson:2009zg,Berti:2009bk,Wei:2010vca,Banados:2010kn,Zaslavskii:2010jd,Zaslavskii:2010aw,Grib:2010dz,Lake:2010bq,Harada:2010yv,Kimura:2010qy,Patil:2011yb,Abdujabbarov:2013qka,Tsukamoto:2013dna,Toshmatov:2014qja,Tsukamoto:2014swa,Armaza:2015eha}. 

 In general relativity, the equation of motion for a spinning test particle is totally different from a test particle, and the effect of spin is non-trivial. Hence, the effect of spin has also been
  also discussed just  for collisions between particles impinging from infinity\cite{maeda2018, Zhang:2018,PhysRevD.98.044006}.
In our previous paper  \cite{maeda2018}(paper I), we 
considered the collision between spinning test particles impinging from infinity and discussed
 the energy efficiency for various processes: the elastic collision, Compton scattering, and inverse Compton scattering.
 In the cases of the elastic collision and Compton scattering, we
found that those extraction efficiencies 
 give twice as larger as the cases of the collision of non-spinning particles.
 
However, we may have to tune the initial conditions for such collisions to be possible.
In order to make the collisional process more realistic, 
we may discuss a collision between a particle on the  innermost stable circular orbit (ISCO) 
and a particle impinging from infinity.
Although  it was shown that the center of mass energy diverges when the collision with
 a particle on the ISCO in the extreme Kerr black hole\cite{Harada:2010yv},
 the efficiency has not been so far calculated in the collision with a particle on the ISCO.
We then consider the collisional Penrose process between a spinning particle on its ISCO and a particle impinging from infinity and calculate the energy efficiency of the ejected particle. 
This is the second paper of a series of papers which provides 
 the maximal efficiency of the collisional Penrose process with spinning particles.

In Sec. II, we briefly summarize the properties of a spinning particle, especially of the ISCO particle, in the extreme Kerr spacetime, and provide the timelike condition of the orbit.
In Sec. III, we study the collision in the extreme Kerr geometry between a particle on the ISCO
and a spinning particle plunging from the radial infinity and analyze the maximal efficiency. In order
We also discuss the collision of one spinning massive particle and one massless particle
(the photoemission process).
Section IV is devoted to concluding remarks.
Throughout this paper,
we use the geometrical units of $c=G=1$ and follow \cite{misner1971gravitation}
for the notations.

\section{BASIC EQUATIONS}
\subsection{Equations of motion  of a spinning particle}
First, we briefly summarize the equation of motion of a spinning particle moving 
on the equatorial plane in the Kerr spacetime.
We consider only the particle motion on the equatorial plane in an extreme Kerr black hole\cite{Saijo:1998mn}.

Introducing the tetrad basis 
\[
e_\mu^{~(a)}  = \left(
    \begin{array}{cccc}
      \frac{r-M}{r} & 0 & 0& -\frac{M(r-M)}{r}\\
      0 &\frac{r}{r-M} & 0& 0\\
      0& 0 & r& 0\\
    -\frac{M}{r-M} & 0 & 0& \frac{r^2+M^2}{r-M}
    \end{array}
  \right)
\,,
\]
we discuss the equation of motion by use of the tetrad components,
which are described by the latin indecies with a bracket.

We define a specific spin vector $s^{(a)}$ by
\beann
	s^{(a)}=-{1\over 2\mu}\epsilon^{(a)}_{~~~(b)(c)(d)}u^{(b)}S^{(c)(d)}
\,,
\enann
where $\mu$ is the mass of a spinning particle, 
$u^{(a)}$ is the specific momentum defined by   $u^{(a)}:=p^{(a)}/\mu$
(the 4 momentum $p^{(a)}$ divided by the mass of a spinning particle $\mu$), and 
$S^{(a)(b)}$ is the spin tensor.
$\epsilon_{(a)(b)(c)(d)}$  is the totally antisymmetric tensor with 
$\epsilon_{(0)(1)(2)(3)}=1$.

For a spinning particle to move just on the equatorial plane, 
 the direction of spin should be perpendicular to the equatorial plane.  
Hence we find only one component of $s^{(a)}$ is non-trivial, i.e., 
\beann
	s^{(2)}=-s.
\enann
When  $s>0$, the particle spin is parallel to the direction of the black hole rotation, 
while it is anti-parallel if $s<0$.
From the supplementary condition; $S^{(a)(b)}p_{(b)}=0$,
 which  fixes the center of mass of a spinning particle, the spin tensor is described as
\beann
	S^{(0)(1)}=- s p^{(3)}\,,~~S^{(0)(3)}= s p^{(1)}\,,~~S^{(1)(3)}= s p^{(0)}.
\enann

For a Killing vector $\xi_{(a)}$, 
we find a conserved quantity of a spinning particle 
\beann
 Q_\xi=p^{(a)}\xi_{(a)}
 +{1\over 2}S^{{(a)}{(b)}} 
 \left( w_{(a)(b)(c)}\xi^{(c)}+e^{\mu}_{~~(a)}\xi_{(b),\mu} \right),
\enann
where Ricci rotation coefficient is defined by 
$w_{(a)(b)(c)}:=e_{(c)\mu;\nu}e^{\mu}_{(b)}e^{\nu}_{(a)}$.
Since there are two Killing vectors in the present spacetime,
we have two conserved quantities; 
the energy $E$ 
and the total angular momentum $J$ 
along the world line of a spinning particle as follows;
\bea
E&=&{r-M \over r}p^{(0)}+{M(r+s)\over r^2}p^{(3)},
\label{Eu0u3}
\\
J&=&
{r-M\over r}(M+s)p^{(0)}+{r(r^2+M^2)+Ms (r+M) \over r^2}p^{(3)}.
\label{Ju0u3}
\notag
\\
\ena
In what follows, we shall use  the normalized variables as
\beann
\tilde E={E\over \mu}\,,~~\tilde J={J\over \mu M}\,,~~\tilde s={s\over  M}
\,,
\enann
\beann
\tilde t={t\over M}\,,~~\tilde r={r\over M}\,,~~\tilde \tau={\tau\over M},
\enann
where $\tau$ is the proper time.
We will drop the tilde just for brevity.

From Eqs. (\ref{Eu0u3}) and (\ref{Ju0u3}), we find
\bea
u^{(0)}&=&
\frac{[r^3+(1+s)r+s]E-(r+s)J }
{ r^2(r-1)\left(1-{s^2\over r^3}\right) }
\label{u0EJ}
\\
u^{(3)}&=&
\frac{J-(1+s)E }
{ r\left(1-{s^2 \over r^3}\right) }
\,.
\label{u3EJ}
\ena
Since we have the normalization condition $u_{(a)} u^{(a)}=-1$,  the radial component of the specific momentum is given by 
\beann
u^{(1)}&=&\sigma \sqrt{(u^{(0)})^2-(u^{(3)})^2-1},
\enann
where $\sigma=\pm 1$ correspond to the outgoing and ingoing motions, respectively.

We should note that the 4-velocity $v^{(a)}=dz^{(a)}/d\tau$, in which $z(\tau)$ is an orbit of a particle, is not always parallel to the specific 4-momentum $u^{(a)}$. 
When we normalize the affine parameter $\tau$
as
\bea
u^{(a)} v_{(a)}=-1\,,
\label{normalization}
\ena
the difference between $v^{(a)}$ and $u^{(a)}$ is given by
\bea
v^{(a)}-u^{(a)}={S^{(a)(b)}R_{(b)(c)(d)(e)}u^{(c)} S^{(d)(e)}
\over 2(\mu^2+{1\over 4}R_{(b)(c)(d)(e)}S^{(b)(c)}S^{(d)(e)})}.
\label{velocity_momentum}
\ena
For the present setting,
this relation between the 4-velocity and the specific 4-momentum is reduced to
\beann
	v^{(0)}={\Lambda}_s^{-1} u^{(0)}, 
	\quad
	v^{(1)}={\Lambda}_s^{-1}  u^{(1)},
	\quad
	v^{(3)}={\left(1+{2s^2\over r^3}\right)\over 
	\left(1-{s^2\over r^3}\right)}  {\Lambda}_s^{-1} u^{(3)}\,,
\enann
where
\beann
  \Sigma_s=
  r^2\left(1-{  s^2\over   r^3}\right),
\quad
 \Lambda_s
=1-\frac{3s^2r[J-(1+s)E]^2 }{{\Sigma}^3_s}.
\enann

Hence, we finally obtain the equations of motion of the spinning particle  as
\beann
 \Sigma_s {\Lambda}_s{d t\over d \tau}
&=& \left(1+{3 s^2\over  r \Sigma_s}\right)
[ J-(1+ s) E]+{ r^2+1\over (r -1)^2 }  P_s
\,,
\\
  \Sigma_s  {\Lambda}_s {d r\over d  \tau}
&=&\pm \sqrt{R_s}
\,,
\\
 \Sigma_s {\Lambda}_s  {d\phi\over d \tau}
&=& \left(1+{3s^2\over  r\Sigma_s}\right)
[ J-(1+s) E]+{1 \over (r -1)^2}  P_s
\,.
\enann
where
\beann
&&
 P_s
=\left[r^2+1+{s\over r}(r+1)\right]  E-\left(1+{ s\over  r}\right) J\,,
\\
&&
R_s
=P_s^2-(r-1)^2
\left[{\Sigma_s^2\over  r^2}+\left[-(1+ s) E+J\right]^2\right]\,.
\enann
By using ${\Sigma}_s$ and ${R}_s$, the radial component of the specific momentum is written by
\bea
u^{(1)}=\sigma{r\sqrt{{R}_s} \over (r-1) {\Sigma}_s }.
\label{specific_radial_momentum}
\ena

\subsection{The innermost stable circular orbit }
Based on the analysis in \cite{Suzuki:1998}, we consider the ISCO of a spinning particle; there are some analyses of this orbit in \cite{PhysRevD.96.064051,PhysRevD.97.084056}.
Since the orbit of a particle on the ISCO is circular, 
$v^{(1)}$ vanishes and then so does $u^{(1)}$.
We then find a constraint on the conserved quantities $E$ and $J$ from Eqs. (\ref{u0EJ}) and (\ref{u3EJ})  as follows:
 We first write Eq.(\ref{specific_radial_momentum}) as
\bea
(u^{(1)})^2=A(E-U_{(+)})(E-U_{(-)})
\ena
where 
\beann
A(r,s)&=&\f{r^2 B}{(r-1)^2 \Sigma_s^2 },
\\
B(r,s)&=&\left[ r^2+1+ s\Big(1+\f{1}{r}\Big) \right]^2-(r-1)^2(1+s)^2, 
\enann
and
\bea
U_{(\pm)}(r,J,s)=XJ \pm \sqrt{(X^2-Y)J^2-Z},
\ena
with 
\beann
X(r,s)&=&\f{1}{B}
\Big[ 
	\Big\{ r^2+1+s\Big( 1+\f{1}{r} \Big) \Big\}
	\Big(1+\f{s}{r} \Big)-(r-1)^2(1+s) 
\Big],
\\
Y(r,s)&=&\f{1}{B}\left[ \Big( 1+\f{s}{r} \Big)^2-(r-1)^2 \right],
\\
Z(r,s)&=&-\f{1}{B}\f{(r-1)^2 \Sigma_s^2}{r^2}. 
\enann

We can regard $U_{(\pm)}(r, J ,s)$ as the effective potential of a particle on the equatorial plane.
Since $U_{(-)}(r)$  usually does not have an extremum and is much less than unity,
we shall consider only $U_{(+)}(r)$.
For a circular orbit with the radius $r=r_0$, the following conditions must be satisfied:
\beann
U_{(+)}(r_0)=E\,,{\rm and}~~ \f{dU_{(+)}}{dr} ({r_0})=0
\,.
\enann
In addition, for the stability of the orbit, we have to impose
\beann
 \f{d^2 U_{+}}{dr^2} (r_0)> 0
\,.
\enann

Since the ISCO is the inner boundary of  stable circular orbits, 
it must satisfy 
\bea
U_{(+)}(r_{\rm ISCO})=E\,, 
\quad
\f{dU_{(+)}}{dr} (r_{\rm ISCO})=0\,,
\quad
\f{d^2 U_{+}}{dr^2}(r_{\rm ISCO})=0
\,,
\label{ISCOcondition}
\ena
where $r_{\rm ISCO}$ is the  ISCO radius.
From the three conditions in Eq. (\ref{ISCOcondition}),
we find the energy $E_{\rm ISCO}$ and angular momentum $J_{\rm ISCO}$ of 
the particle on the ISCO, and the ISCO radius $r_{\rm ISCO}$ in terms of the spin $s$.
In the extreme Kerr spacetime,
we find  $r_{\rm ISCO}=1$ (the horizon radius),
and then 
\bea
E_{\rm ISCO}&=&\f{(1-s^2)}{\sqrt{3(1+2s)}}
\,,
\\
J_{\rm ISCO}&=&2E_{\rm ISCO}
\label{spinnning_angular_momentum_ISCO}
\,.
\ena 
From Eq. (\ref{velocity_momentum}), 
we find
\beann
v^{(0)}_{\rm ISCO}=\f{u^{(0)}_{\rm ISCO}}{\tilde{\delta}} \left( 1-\frac{s^2}{r^3} \right), 
\quad
v^{(3)}_{\rm ISCO}=\f{u^{(3)}_{\rm ISCO}}{\tilde{\delta}} \left( 1+\frac{2s^2}{r^3} \right), 
\enann
where
\beann
\tilde{\delta}=1-\f{s^2}{r^3}\left[ 1+3(u^{(3)}_{\rm ISCO})^2  \right]
\,.
\enann

\subsection{Constraints on the Orbits}
Here, we stress two important points:
\begin{enumerate}
\item 
In order for a particle to reach  the horizon $r_H:=1$,
the radial function $R_s$ must be non-negative for $r\geq r_H$.
\item
As seen in the previous section, the 4-velocity $v^{(a)}$ is not always parallel to the specific 4-momentum. 
Hence, we have to impose the timelike condition $v^{(a)}v_{(a)}<0$ even though $u^{(a)}u_{(a)}=-1$ is imposed. 
\end{enumerate}
Before discussing the collisional Penrose process,
we must treat these points properly.
Since we have already discussed these points for the particles plunging from infinity in \cite{maeda2018}, 
we shall  discuss only the case for the ISCO particle.
For the first condition, 
$R_s(r_{\rm ISCO})=0$ always holds by definition, so it is always satisfied. 

As for the timelike condition $v^{(a)}v_{(a)}<0$,
we find 
\bea
 E_{\rm ISCO}^2<{(1-s)^2(1+s)^4\over 3 s^{2} (2+s^2) }\,.
\label{Es_constraint}
\ena
 Since the conserved energy $E_{\rm ISCO}$ is a function of the spin $s$, 
this inequality is reduced to 
\beann
(1-s^2)^2 (s^4-2s^3-3s^2-4s-1)<0.
\enann
From this inequality, we obtain the constraint of the spin $s$ for the ISCO particle,
 i.e.,
\bea
s_{\text{min}}^{\text{ISCO}} \le s \le s_{\text{max}}^{\text{ISCO}}
\,,
\ena 
where $s_{\text{min}}^{\text{ISCO}} \approx -0.302776$ and 
$s_{\text{max}}^{\text{ISCO}}=1$.
The  energy of the ISCO particle is bounded as 
\beann
0\leq E_{\rm ISCO}\leq E_{\rm ISCO}^{\rm max}\approx 0.8349996
\enann

Note that 
for a particle plunging from infinity, we have the constraint such that
\beann
s_{\rm min}\leq s \leq s_{\rm max}
\,,
\enann
where 
$s_{\rm min}\approx -0.2709$ and $s_{\rm max}\approx 0.4499$ are the solutions of
$
s^6+2s^5-4s^4-4s^3-7s^2+2s+1=0
$
with the condition 
$-1\leq s_{\rm min}<s_{\rm max}\leq 1 $.

\section{ MAXIMAL EFFICIENCY OF COLLISION OF PARTICLES }
\label{collision}
Now, we discuss the collision of two particles 1 and 2, 
whose 4-momenta are $p_1^{(a)}$ and $p_2^{(a)}$.
We then assume that 
particle 1 is on the ISCO, while 
particle 2 is impinging from infinity.
In Appendix \ref{appendix_bsw}, we show that 
the center of mass energy can take arbitrary value in the collision between those two spinning particles just as the usual BSW effect.
We will then analyze the maximal efficiency  
of the energy extraction, i.e., 
how much energy we can extract from an extreme black hole. 
For this purpose,
we shall 
follow the same procedure as in paper I.
The difference is that particle 1 is moving on the ISCO orbit with the radius 
$r_{\rm ISCO}$.
In the extreme Kerr spacetime, the ISCO radius is $r_{\rm ISCO}=1$.
Hence, the collision must take place very close to the horizon ($r_H=1$). 
We then assume that the collisional point  is 
given by $r_c=1/(1-\epsilon)$ ($0<\epsilon \ll 1$).

At the collisional point $r_c$, we impose the following conservations:
\beann
	E_1+E_2&=&E_3+E_4,
	\\
	J_1+J_2&=&J_3+J_4,
	\\
	s_1+s_2&=&s_3+s_4,
	\\
	p_1^{(1)}+p_2^{(1)}&=&p_3^{(1)}+p_4^{(1)}.
\enann
After the collision, we assume that 
particles 3 with the 4-momentum $p_3^{(a)}$ is going away to infinity, 
while particle 4 with the 4-momentum $p_4^{(a)}$ falls into the black hole.
In order for particle 2 to reach the horizon from infinity,
particle 2 must be subcritical ($J_2<2E_2$) and ingoing ($\sigma_2=-1$).
So we assume that $\sigma_4=\sigma_2=-1$.

We then expand the radial component of the 4-momentum $p^{(1)}$ in terms of $\epsilon$ 
as
\beann
p^{(1)}\approx \sigma {|2E-J|\over \epsilon(1-s)}+\cdots
\,.
\enann
The conservation of the radial components  of the momenta
($p_1^{(1)}+p_2^{(1)}=p_3^{(1)}+p_4^{(1)}$) yields
\bea
	- {|2E_2-J_2|\over 1-s_2}
	&=&\sigma_3{|2E_3-J_3|\over 1-s_3} -{|2E_4-J_4|\over 1-s_4}
	+O(\epsilon),
	\notag \\
	\label{leading_order}
\ena
in which we use $J_1=2E_1$.  
Here, we also assume $s_2=s_4$ for simplicity.
This means that particle 1 becomes particle 3 without change of the spin angular momentum after the collision.
Then, the above conservation is written as
\beann
	\left[ 
	\sigma_3{\text{sign} [2 E_3-J_3] \over 1- s_3} +{1\over 1-s_4}
	\right] (2E_3-J_3)=O(\epsilon).
\enann

The above setting gives 
\bea
	J_3&=&2 E_3 (1+\alpha_3\epsilon+\beta_3 \epsilon^2+\cdots),
	\label{case3_1}
\ena
where $\alpha_3$  and $\beta_3$ are parameters of $O(\epsilon^0)$.
Since particle 2 is subcritical ($J_2<2E_2$), 
the angular momentum $J_2$ is written as
\bea
	J_2&=&2 E_2 (1+\zeta),
	\label{case3_2}
\ena
where $\zeta<0$ with  $\zeta=O(\epsilon^0)$.
From the conservation laws, 
we find 
\bea
	E_4=E_1+E_2-E_3\,,~~~J_4=J_1+J_2-J_3,
	\label{conservation}
\ena
giving 
\beann
	J_4=2 E_4\left(1+{E_2\over E_4}\zeta+\cdots\right).
\enann

Hereafter, we consider two cases: 
\begin{description}
\item[[A]] collision of two massive particles (MMM), and 
\item [[B]] collision of massless and massive particles (MPM),
\end{description}
where we use the symbols MMM and MPM following 
\cite{Leiderschneider:2015kwa}.
In the case of {[A]} (MMM), 
we assume all the masses of the particles are the same, i.e.
$\mu_1=\mu_2=\mu_3=\mu_4=\mu$.
In the case of {[B]} (MPM), particles 2 and 4 are massless and non-spinning,
because a photon has no stable circular orbit,  
while particles 1 and 3  are massive with the same mass, i.e., $\mu_1=\mu_3=\mu$.
From the conservation of spin, we obtain $s_1=s_3$.
Since the massive particle is ejected by the incoming photon, 
we shall call process {[B]} (MPM) photoemission.
Note that in the first paper, we called the process the inverse Compton scattering because 
a massive particle with large energy is going out after the collision.

Next, we evaluate  $E_2$ and $E_3$ for the cases [A] and [B] separately.

\subsection{Maximal Efficiency in Case [A] (MMM: Collision of two massive particles)}
For the massive particle, the radial component of the specific 4-momentum 
is given as
\bea
&&u^{(1)}=\sigma {r\sqrt{R_s}\over \Sigma_s\sqrt{\Delta}}
\notag \\
&&
={\sigma\sqrt{r^2\left[(r^3+(1+s)r+s)E-(r+s)J\right]^2-(r-1)^2\left[(r^3-s^2)^2+r^4(J-(1+s)E)^2\right]}
\over (r-1)(r^3-s^2)}
\,.
\notag \\ 
\label{radial_momentum}
\ena
By expanding $u^{(1)}$ with the conditions in Eqs. (\ref{case3_1}) and (\ref{case3_2}) in terms of $\epsilon$, 
we find Eqs. (3.15)-(3.17) in paper I.
Note that for particle 1 on the ISCO, $u_1^{(1)}=0$.

Since $u_1^{(1)}+u_2^{(1)}=u_3^{(1)}+u_4^{(1)}$, 
we find the leading order of $\epsilon^{-1}$ is trivial.
From the next leading order, i.e., $O(\epsilon^0)$, we find
\beann
&&
\sigma_3{f(s_1,E_3,\alpha_3)\over 1-s_1^2}
={\left[E_1 (2+s_2) -E_3 g_1(s_2,\alpha_3)
\right]\over 1-s_2^2}
\,,
\enann
where 
\beann
f(s,E,\alpha)&:=&\sqrt{E^2 [3 - 2 \alpha (1+s)][1 + 2 s - 2 \alpha (1+s)] - (1 -  s^2)^2}
\,,
\\
g_1(s,\alpha)&:=&2+s-2\alpha(1+s)
\,.
\enann
This equation is reduced to
\bea
{\cal A}E_3^2-2{\cal B}E_3+{\cal C}=0
\,,
\label{eq_E3}
\ena
where
\bea
{\cal A}&=&
-[3-2\alpha_3(1+s_1)][1+2s_1-2\alpha_3(1+s_1)]
+\f{(1 - s_2^2)^2} {(1-s_1^2)^2} g_1^2(s_2,\alpha_3)
\label{calA}
\\
{\cal B}&=&g_1(s_2,\alpha_3)\f{(1 - s_2^2)^2} {(1-s_1^2)^2} (2+s_2)E_1
\label{calB}
\\
{\cal C}&=&
\f{(1 - s_2^2)^2} {(1-s_1^2)^2}(2+s_2)^2 E_1^2+ (1-s_1^2)^2
\label{calC}
\,,
\ena
with the condition such that 
$E_3\leq E_{3, {\rm cr}}$ for $\sigma_3=1$, or  
$E_3\geq E_{3, {\rm cr}}$ for $\sigma_3=-1$,
where
\beann
E_{3, {\rm cr}}:={2+s_2 \over g_1(s_2,\alpha_3)}E_1
\,.
\enann

In the case of $\sigma_3=1$, we do not expect large efficiency
since the energy $E_3$ has the upper bound $E_{3, {\rm cr}}$,
which magnitude is the order of $E_1$.
In fact, we find the efficiency for the case of $\sigma_3=1$ is not so high in Appendix \ref{appendix}. 
Hence, we will focus on the case of $\sigma_3=-1$, i.e., 
particle 3 is assumed to be ingoing after the collision.
For particle 3 to go back to infinity, the orbit must be supercritical 
($J_3>2 E_3$), which means either $\alpha_3 >0$ or $\alpha_3=0$ with $\beta_3>0$.

From Eq.(\ref{calA}), we obtain the larger output energy $E_3$
in terms of  $s_1$, $s_2$ and $\alpha_3$:
\bea
E_3&=&E_{3,+}:={{\cal B}+\sqrt{{\cal B}^2-{\cal A}{\cal C}} \over {\cal A}}
\label{energy_E3}
\,.
\ena

From the next leading order terms, we obtain 
\bea
{\cal P}E_2
   = (1 - s_2)^3 (E_3-E_1)^2
   \,,
\label{energy_E2}
\ena
where
\bea
{\cal P}:&=& 2 (E_3-E_1)(1-s_2)^3 
\notag \\ &&\quad
+ 4\zeta\Big\{ 2(1+s_2)E_3[ \alpha_3 (2 +s_2)- \beta_3 (1 - s_2^2) ] 
-s_2(2+s_2)^2 (E_3-E_1 )
\notag \\ &&\quad
-\sigma_3 \f{(1 - s_2^2)^2} {(1-s_1^2)^2} \Big{[}{ E_3^2\over f(s_1,E_3,\alpha_3)}
 \Big(h(s_1) - 2 (1 + s_1)^2(2 + s_1)g_2(s_1,\alpha_3)
\notag \\ &&\quad
 +2\beta_3 (1+s_1)  (1 - s_1^2)g_1(s_1,\alpha_3)\Big)
\Big{]}
\Big\}
\,,~~~
\label{calP}
\ena
in which 
\beann
g_2(s,\alpha)&:=&\alpha (2 + s-2  \alpha)
\enann

Since this fixes the value of  $E_2$, we obtain
the efficiency  by 
\beann
\eta={E_3\over E_1+E_2}
\,,
\enann
when  $\alpha_3, \beta_3$ and $\zeta$ are given.

\subsubsection{Non-spinning particles}
We first consider the collision of non-spinning particles: $s_1=s_2=0$. 
In this case, the energy of particle 1 ($E_1$), is $1/\sqrt{3}$.
Since particle 2 is plunging from infinity, we have the constraint of $E_2\geq 1$ for a massive particle. 
Hence, in order to obtain the maximal efficiency, 
we need to obtain the maximal energy of particle 3 and the minimal energy of  particle 2, i.e., $E_2=1$.

\begin{figure}[h]
	\centering\includegraphics[width=8cm]{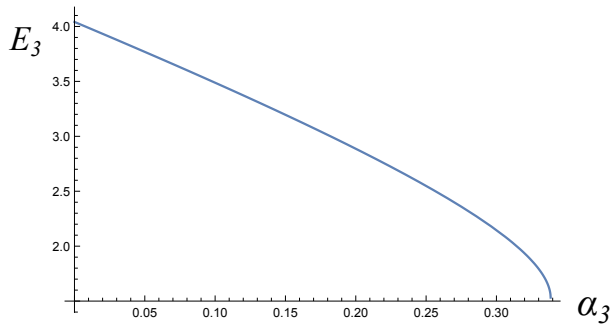}
\caption{
The energy of particle 3, $E_3$ in terms of $\alpha_3$. 
The maximal value of this is given at $\alpha_3=0$.
} 
\label{E3_spinless_elastic_isco}
\end{figure}

The energy of particle 3 is given by
\beann
E_3=\frac{1}{\sqrt{3}}
	\Big[
	4(1-\alpha_3)
	+\sqrt{4(3-2\alpha_3)(1-2\alpha_3)-3}
	\Big].
\enann
Then we find that from Fig.{\ref{E3_spinless_elastic_isco}}, 
the maximal value of the energy of particle 3 is given at $\alpha_3=0+$. 

Next, when $\alpha_3=0+$, the energy of particle 2 is given by
\bea
E_2=\frac{36 \sqrt{3}}{ 36+7 \zeta (4 \beta_3+7) },
\label{E2_spinless}
\ena
which becomes unity when $\beta_3$ is taken to be 
\beann
\beta_3=\frac{1}{4}\Big( \frac{36(\sqrt{3}-1)}{7\zeta}-7 \Big).
\enann
As a result, we obtain the maximal efficiency for the collision of 
non-spinning particles, $\eta_{\rm{max}}=7(\sqrt{3}-1)/2 \approx 2.562$.

\subsubsection{Spinning particles}
Next, we consider the collision of spinning particles.
In our approach, the energy of particle 1 ($E_1$) is a function of $s_1$ while the energy of particle 2 ($E_2$) is a function of $s_1$, $s_2$, $\alpha_3$, $\beta_3$ and $\zeta$, which are the orbit parameters of particles 2 and 3. The energy of particle 3 ($E_3$) is a function of $s_1$, $s_2$ and $\alpha_3$.
When particle 1 comes from infinity as in the previous paper, 
we can look $E_1$ as a parameter and treat $E_1$ and $E_3$ separately. 
However, in the current setup, we must modify how to derive the maximal efficiency since $E_1$ is a function of $s_1$. This is the main difference from the analysis in the previous paper.

When the energy of particle 1 ($E_1$) is fixed for a given value of $s_1$, 
it is clear that the larger efficiency is obtained for larger $E_3$ as well as smaller $E_2$.
Hence, we expect that the efficiency for an elastic scattering could take the maximum value as $E_3/(E_1+1)$, which depends on $s_1$, $s_2$, and $\alpha_3$ because we have the constraint for particle 2, $E_2\geq 1$,
as for the non-spinninig particles. 
In order to justify this expectation, 
it is necessary to show that $E_2=1$ is possible for some choices of the remaining parameters ($\zeta$ and $\beta_3$)
after finding $s_1$, $s_2$, and $\alpha_3$ giving the maximal efficiency of $E_3/(E_1+1)$.

As for the energy of particle 3 ($E_3$),
since the orbit of particle 3 is near critical, 
in order to determine $E_3$ from Eq. (\ref{energy_E3}) properly,
we have to consider two constraints:
$E_3\geq E_{3, {\rm cr}}$ for $\sigma_3=-1$
and the timelike condition (\ref{Es_constraint}).
Due to the timelike condition, 
in order to find the large value of $E_3$, 
the spin magnitude $s_3(=s_1)$ must be small.
For a small value of $s_1$, we 
 find $\alpha_3 \approx 0$ gives the largest efficiency of $E_3/(E_1+1)$.

\begin{figure}[h]
	\centering\includegraphics[width=8cm]{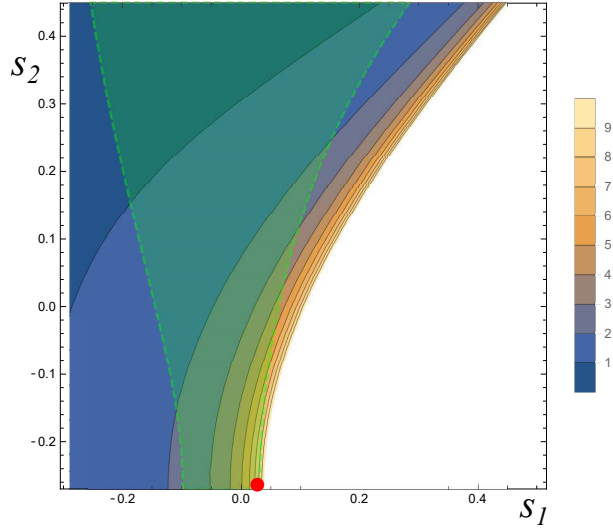}
\caption{
The contour map of $E_3/(E_1+1)$ in terms of $s_1$ and $s_2$ when we set $\alpha_3=0+$.
In the light green region, the timelike condition for the particle 3 orbit is satisfied.
The maximal value of $E_3/(E_1+1)\approx 8.442$ is obtained when $s_2=s_{\rm min}\approx-0.2709$ and $s_1\approx 0.03196$ (the red point in the figure).} 
\label{maxE3_E1_isco}
\end{figure}

Hence, setting $\alpha_3=0+$, we analyze the maximal efficiency
in terms of $s_1$ and $s_2$.
In Fig.\ref{maxE3_E1_isco}, we show a contour map of $E_3/(E_1+1)$ in terms of $s_1$ and $s_2$. Form this figure, we find that the red point, which is $(s_1, s_2)\approx (0.03196, s_{\rm min})$,
 gives the maximal value of $E_3/(E_1+1)$.

Then, we have to confirm that  $E_2=1$ is possible 
for certain values of remaining parameters ($\zeta$ and $\beta_3$). 
\begin{figure}[h]
	\centering\includegraphics[width=7cm]{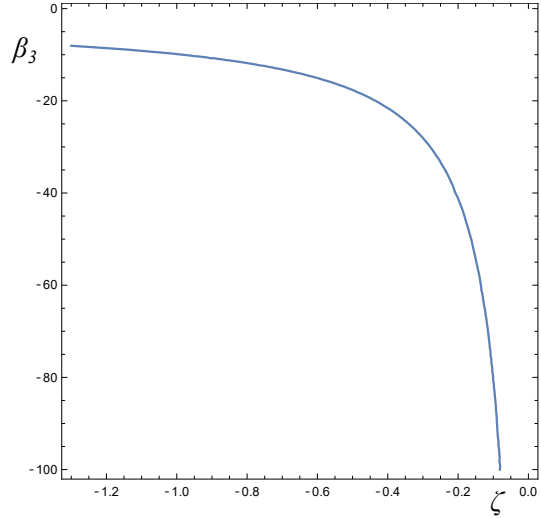}
\caption{
The relation giving $E_2=1$ between $\zeta$ and $\beta_3$.
The other parameters ($s_1$, $s_2$ and $\alpha_3$) are chosen so as to give the maximal value of $E_3/(E_1 + 1)$.
The timelike condition for the particle 2 orbit
gives the constraint of $\zeta_{\rm min}<\zeta<0$
with $\zeta_{\rm min}\approx -1.271$.
} 
\label{E2_isco}
\end{figure}
If the condition, $E_2=1$, is satisfied,
we obtain the relation between $\zeta$ and $\beta_3$ from Eq. (\ref{energy_E2}), which is a linear equation of $\beta_3$. 
As for $\zeta$, since particle 2 also has spin $s_2$, we have to consider the timelike condition for the particle 2 orbit. 
This gives the constraint on $\zeta$, 
\beann
\zeta_{\rm min}<\zeta
<0
\,,
\label{timelike_E2}
\enann
where 
\beann
\zeta_{\rm min}:= -{(1-s_2)\over 2}\left[1+{(1-s_2)(1+s_2)^2\over \sqrt{3s_2^2(2+s_2^2)}}\right]
\,.
\enann
For the parameters giving the maximal value of $E_3/(E_1+1)$, 
we obtain $\zeta_{\rm min}\approx -1.271$.
Under this constraint,
we find in Fig. \ref{E2_isco} the relation giving the condition $E_2=1$
between $\zeta$ and $\beta_3$.

Thus, we find the maximal efficiency is given by 
\bea
\eta_{\rm max}=\f{E_3}{E_1+1}\approx 8.442.
\ena
\subsection{Maximal Efficiency in Case {[B]} (MPM: photoemission)}
For the collision of a massless particle (photon) and a massive particle, 
we should assume that particle 1 (the ISCO particle) is massive 
and the incoming particle is massless because there is no ISCO 
for massless particles.
Hence we consider the photoemission process, i.e., a massive particle is emitted via a collision process by an incoming photon.

For the momenta of the massive particles 1 and 3, 
 the radial componponents of 4-momentua do not change,
 while for the massless particles 2 and 4, we find
\bea
p_2^{(1)}&=&2\epsilon^{-1}E_2\zeta -2E_2(1+2\zeta)
-\epsilon{E_2 (1-4\zeta^2)\over  4\zeta}+O(\epsilon^2),
\label{photon2}
\\
p_4^{(1)}&=&2\epsilon^{-1}E_2\zeta-2\left[E_4+2E_2\zeta+E_3\alpha_3\right]
-\epsilon{E_4^2-8E_2E_3(2\alpha_3-\beta_3)\zeta
-4E_2^2\zeta^2\over 4 E_2\zeta}
+O(\epsilon^2) 
\,.
\notag \\
\label{photon4}
\ena
where $E_4=E_1+E_2-E_3$

From the conservation of the radial components of the 4-momenta, 
we find 
\bea
E_3&=&
\left. 
{{\cal B}+\sqrt{{\cal B}^2-{\cal A}{\cal C}} \over {\cal A}}
\right|_{s_2=0}
\label{Inverse_Compton_energy_E3}
\,,
\ena
and 
\bea
E_2&=&
   \left.{(E_3-E_1)^2 \over {\cal P}} \right|_{s_2=0}
   \,,
\label{Inverse_Compton_energy_E2}
\ena
where ${\cal A}, {\cal B}, {\cal C}$ and ${\cal P}$ are given by Eqs.
 (\ref{calA}),  (\ref{calB}),  (\ref{calC}) and (\ref{calP}),
 which is evaluated with $s_2=0$.
Note that the representations for $E_2$ and $E_3$, i.e. Eqs.(\ref{Inverse_Compton_energy_E3}) and (\ref{Inverse_Compton_energy_E2}), respectively coincide with Eqs.(\ref{energy_E3}) and (\ref{energy_E2}) for the  collision of a spinning massive particle and a non-spinning massive particle.

\subsubsection{Non-spinning particles}
We consider the collision between a non-spinning massive particle and s massless particle: $s_1=s_2=0$.
As the non-spinning particles in the elastic collision, 
the energy of particle 1 ($E_1$) is given by $1/\sqrt{3}$ and
the maximal value of the energy of particle 3 is given at $\alpha_3=0+$.
On the other hand, the energy of particle 2 is given by Eq.(\ref{E2_spinless}).
Since particles 2 plunges from infinity,
we have the constraint for the energy of particle 2, $E_2\ge 0$.
When we take the limit: $\zeta \beta_3 \rightarrow \infty$ in Eq.(\ref{E2_spinless}), the minimal energy of particle 2 $E_2 \rightarrow 0$ is obtained. Thus, we obtain the maximal efficiency as $\eta_{\rm{max}}=7$.

\subsubsection{Spinning particle + massless particle}
As describe before, when the energy of particle $E_1$ is fixed, the larger efficiency is obtained for larger $E_3$ as well as smaller $E_2$.
Hence, we first discuss $E_3/E_1$ which depend on $\alpha_3$ and $s_1$ because we cannot treat $E_1$ and $E_3$ separately and have the constraint $E_2\ge 0$ for a massless particle.
After finding these parameters which give the maximum value of $E_3/E_1$, we need to confirm that the condition $E_2\rightarrow0$ is possible for the remaining parameters ($\zeta$ and $\beta_3$). If the condition $E_2 \rightarrow 0$ is possible for the remaining parameters, the maximal efficiency is really given by $\eta_{\rm max}=E_3/E_1$.

\begin{figure}[h]
	\centering\includegraphics[width=8cm]{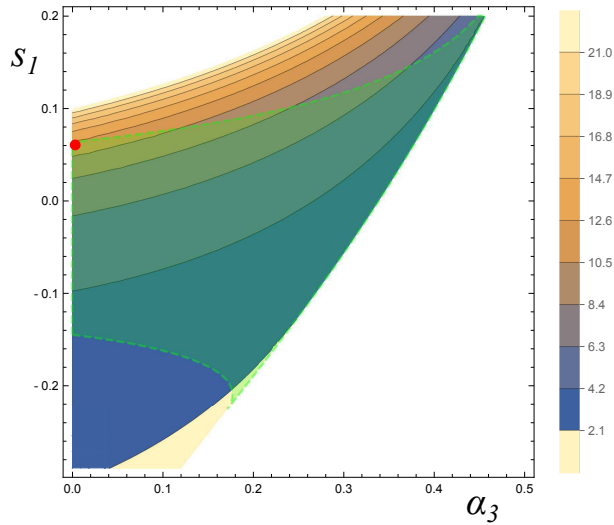}
\caption{The contour map of ${E_3/E_1}$ in terms of $\alpha_3$ and $s_1$.
In the light-green region, the timelike condition for particle 3 is satisfied.
The maximum value of $E_3/E_1=12.54$ is obtained 
at the red point $(\alpha_3, s_1)=(0+, 0.06360)$.}
\label{Inverse_Compton_eff_isco}
\end{figure}

The energy ratio $E_3/E_1$ depends on only two parameters $\alpha_3$ and $s_1$ and we can directly draw the contour map as in Fig.\ref{Inverse_Compton_eff_isco}. 
From this figure, we find the red point, which is $(\alpha_3,s_1)=(0, 0.06360)$, gives the maximal value of $E_3/E_1$.

Next, we confirm that $E_2\rightarrow 0$ is possible for the remaining parameters ($\zeta$ and $\beta_3$) after we set  $(\alpha_3, s_1)= ( 0+, 0.06360)$ giving the maximal value of $E_3/E_1$. As for the remaining parameters, $\zeta$ is constrained as $-\infty<\zeta<0$ because particle 2 is non-spinning, while $\beta_3$ takes arbitrary value as long as $\alpha_3>0$ is satisfied.
Under these constraints, we find the asymptotic behavior of ${\cal P}$ from Eq. (\ref{calP}) when $\zeta\beta_3\rightarrow \infty$:
\beann
{\cal P}\approx 
8E_3 \zeta \beta_3\left[{E_3(2+s_1)\over (1-s_1)f(s_1, E_3, 0)}-1\right]
\,.
\enann
To take this limit, $\beta_3$ must be negative. 
From this asymptotic representation, it is easy to find $E_2\rightarrow 0$ when $\zeta\beta_3\rightarrow \infty$. 

Thus, we find the maximum efficiency 
$\eta_{\rm max}\approx 12.54$  for the photoemission process.

\section{CONCLUDING REMARKS}
We have studied  the collisional Penrose process for non-spinning and spinning particles 
around an extreme Kerr black hole.
For the collision between a particle on the ISCO orbit and a particle impinging from infinity, we have evaluated the maximal efficiency of the energy extraction. 
We summarize our present result as well as the previous one in the paper I \cite{maeda2018}
 in Table I. 

In the non-spinning case, we find that 
the maximal efficiency is 2.562 for the elastic collision, 
 and 7 for the photoemission process. 
For spinning particles, we obtain 
  the maximal efficiency in the elastic scattering (MMM0)  as $\eta_{\text{max}} \approx 8.442$, and $\eta \approx 12.54$
 for the photoemission process (MPM0).
When we take spin into account, we find that the efficiency 
 becomes larger both in the elastic collision and for the photoemission process.

Note that for the collision between particles impinging from infinity, the maximal efficiency 
becomes the largest in the Compton scattering (PMP+) when the energy of particle 1 ($E_1$) takes $E_1 \rightarrow \infty$. 
This result does not change even if the spin is taken into account.
In the present case, however, particle on the ISCO should be massive, which
means that the PMP process is not possible.

In the current analysis, compared with the non-spinning case, 
the maximal efficiencies for both the elastic scattering and the photoemission process 
 become twice larger than the non-spinning case. We can conclude that spin plays an important role also in the collision with the ISCO particle.
Note that the efficiency does not change significantly in the case of the photoemission process. This is because the absorbed massless particle is non-spinning.

\begin{table}[htbp]
\caption{
    The maximal efficiencies and energies 
    for three collisions between a particle in its ISCO 
    and a particle impinging from infinity. We include 
    the previous result for the non-ISCO orbit case. 
        Following \cite{Leiderschneider:2015kwa}, 
    we use the symbols of MMM0, PMP0, MPM0 for each process,
    where ``0'' means a collision with a particle on the ISCO orbit.
    We also include the cases of three collisions with particles impinging from infinity  
    discussed in paper I \cite{maeda2018} as a reference.
    The maximal efficiencies and maximal energies are always enhanced 
    when the spin effect is taken into account. 
    }
\begin{center}
\scalebox{0.8}{
{%
\begin{tabular}{|c|c||c|c|c|c|}
\hline 
\raisebox{-6pt}{collisional process}&&
	spin &input energy& output energy  &maximal\\ [-0.4em]
	&&$(s_{\rm ISCO} ({\rm or} s_1),s_2)$
	&$(E_{\rm ISCO},E_2)$&($E_3$)&efficiency\\
\hline \hline
&MMM0& non-spinning & $( 0.5773 \mu, \mu )$ & $4.041 \mu$ & $2.562$
\\
\cline{3-6}
 Collision of &ISCO& $ ( 0.03196\mu M, -0.2709 \mu M ) $ & $ ( 0.5591 \mu, \mu ) $ & $13.16 \mu$ &$8.442$ \\ 
\cline{2-6} 
two massive particles &&&&&
\\[-.9em]
\cline{2-6} 
&MMM+& non-spinning &\raisebox{-6pt}{$(\mu,\mu)$}&$12.66 \mu$&$6.328$
\\[-.4em] \cline{3-3}\cline{5-6}
&non-ISCO&  $(0.01379\mu M, -0.2709\mu M)$&&$30.02 \mu$ &$15.01$ \\ [.1em]
 \hline
\hline
Photoemission
&MPM0 &  non-spinning& $( 0.5773 \mu, 0)$ & $ 4.041 \mu$ & $ 7$
\\
\cline{3-6}
&ISCO& $( 0.06360 \mu M, 0)$ & $( 0.5416 \mu, 0)$ &$ 6.791 \mu$ & $ 12.54$
\\
\cline{1-6} 
Compton scattering  &PMP+& non-spinning&\raisebox{-6pt}{$(+\infty,\mu)$}&$+
  \infty$&$13.93$
   \\[-.4em]  \cline{3-3}\cline{5-6}
   &non-ISCO& $(0, -0.2709\mu M)$&&$+\infty$ &$26.85$\\ [.1em]
\cline{1-6} 
Inverse Compton scattering & MPM+ &  non-spinning&\raisebox{-6pt}{$(\mu,0)$}&$12.66 \mu$&$12.66$
 \\[-.4em]  \cline{3-3}\cline{5-6}
&non-ISCO& $(0.02679\mu M, 0)$&&$15.64 \mu$ &$15.64$\\
\hline 
\hline 
\end{tabular}
}
\label{summary1}
}
\end{center}
\end{table}
Our analysis was performed for an extreme Kerr black hole.
Since the existence of an extreme black hole may not be likely \cite{thorne1974disk}, 
we will  extend the present  analysis into the case for a non-extreme black hole.

\section*{ACKNOWLEDGMENTS}
We would like to thank Hirotada Okawa, Shingo Suzuki, and Keigo Shimada for  useful information and comments. 
This work was supported in part by JSPS KAKENHI Grant Numbers   
JP17H06359 (KM) and JP19K03857 (KM). 

~~\\
\appendix

\section{
The BSW effect for the collision with a spinning particle on the ISCO 
}
\label{appendix_bsw}
The center of mass energy $E_{\rm{cm}}$ is defined as
\beann
E_{\rm{cm}}^2=-(p_{1(a)}+p_{2(a)})(p_{1}^{(a)}+p_{2}^{(a)}).
\enann
We consider a collision between spinning particles 
which have the same mass $\mu$.  
If particle 1 is on the ISCO, $p_1^{(1)}=0$ holds. 
Hence, we find
\beann
	\frac{E_{\text{cm}}^2}{2\mu^2}
	&=&
	1-\frac{ r^4 A_s(s_1,E_1,J_1) A_s(s_2,E_2,J_2)}{(r^3-s_1^2)(r^3-s_2^2)}
	+\frac{r^2 B_s(s_1,E_1,J_1)B_s(s_2,E_2,J_2)}{(r-1)^2(r^3-s_1^2)(r^3-s_2^2)}
\enann
where
\beann
A_s(s,E,J)&=&  (1 + s)E - J, \\
B_s(s,E,J)&=& (r^3+ (1+s)r + s) E-J ( r+s) .
\enann
In addition, the relation between the angular momentum $J_1$ and the energy $E_1$
is given by $J_1=2E_1$.
Assuming the collision takes place near the horizon, i.e., $r=1+\epsilon$,
the above equation becomes:
\beann
	&&
	\frac{E_{\text{cm}}^2}{2\mu^2}
	=
	\frac{(2+s_1)}{(1-s_1^2)(1-s_2)}\frac{E_1(2E_2-J_2) }{\epsilon}+O(\epsilon^0).
\enann 
Thus, we find that the center of mass energy $E_{\text{cm}}$ diverges 
at the horizon ($\epsilon\rightarrow 0$).

\section{The case for particle 3 with $\sigma_3=1$}
\label{appendix}
\subsection{Case[A] (MMM: Collision of two massive particles)}
\begin{figure}[h]
	\centering\includegraphics[width=8cm]{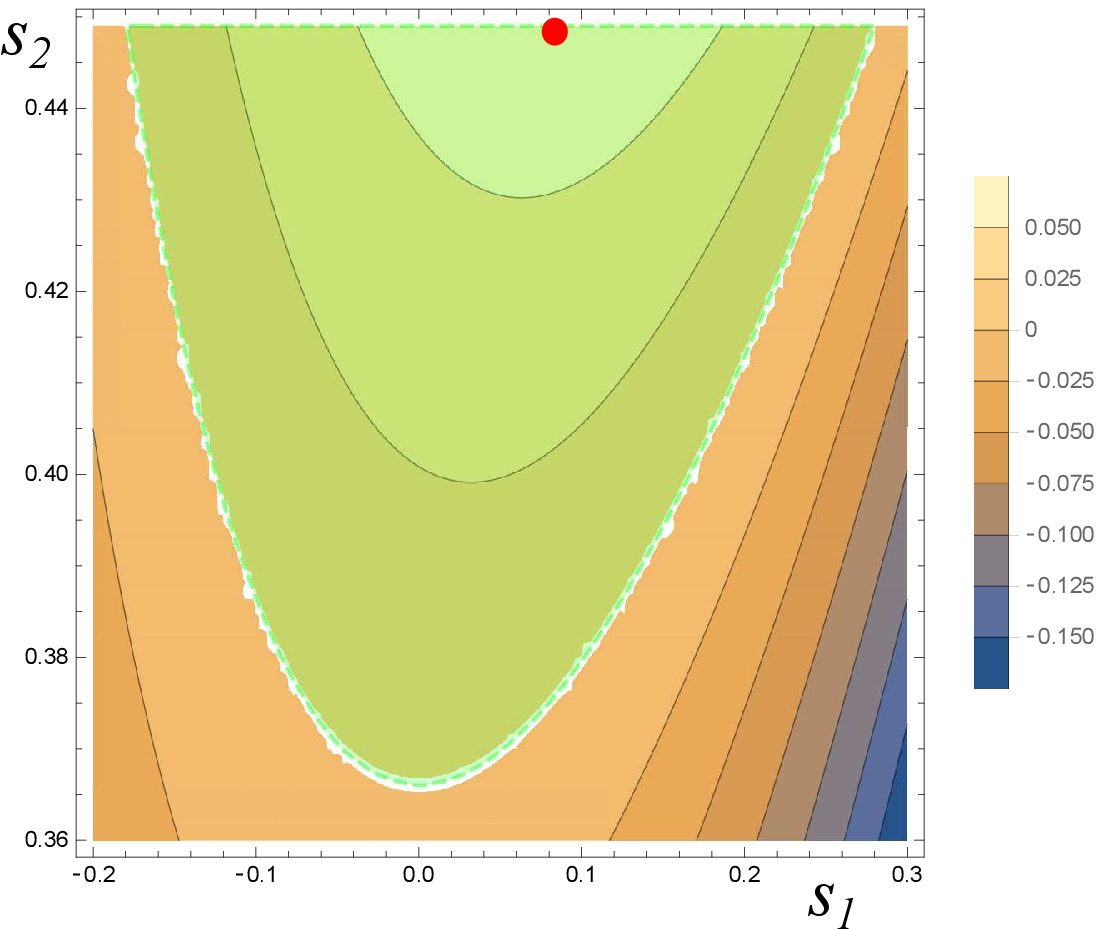}
\caption{The contour map of $E_3/(E_1+1)$ for $\alpha_3=\alpha_3(s_1,s_2)$ in terms of $s_1$ and $s_2$.
The timelike condition of particle 3 is satisfied in the light-green shaded region.
In this case, $E_3/(E_1+1) \approx 0.06587$ is obtained 
as the maximum value at the red point $(s_1,s_2) \approx (0.08230, 0.449)$.
}
\label{Appendix_eff2_Elastic_isco}
\end{figure}

In this case, the condition $E_3 \le E_{3,({\rm cr})}$ must be satisfied.
Hence, the larger root of Eq. (\ref{eq_E3}): $E_{3,+}$ is excluded, and
we obtain the energy of particle 3 ($E_3$) as
\beann
E_3&=&E_{3,-}:={{\cal B}-\sqrt{{\cal B}^2-{\cal A}{\cal C}} \over {\cal A}}
\,.
\enann
From the condition $E_3 \le E_{3,({\rm cr})}$, the energy of particle 3 ($E_{3,-}$) is less than or equal to $E_{3,({\rm cr})}$.
$E_{3,({\rm cr})}$ is a function of $\alpha_3$, $s_1$ and $s_2$ which increases monotonically for $\alpha_3< \alpha_{3, \infty}:={2+s_2 \over 2(1+s_2)}$.
In this region, $E_{3, {\rm cr}}$ is positive and 
$E_{3, {\rm cr}} \rightarrow \infty$ as $\alpha_{3}\rightarrow \alpha_{3,\infty}$. 
For $\alpha_3 > \alpha_{3, \infty}$, $E_{3, {\rm cr}}$ becomes negative and this case should be excluded.
As $\alpha_3$ increases in $0 < \alpha_3< \alpha_{3, \infty}$, the energy of particle 3 ($E_{3,-}$) also increases but faster than $E_{3, {\rm cr}}$ and reaches the upper bound $E_{3, {\rm cr}}$ at some value of $\alpha_3=\alpha_{3,{\rm cr}}$.
Hence, for given values of $s_1$ and $s_2$, it is sufficient to find $\alpha_3$ satisfying $E_3=E_{3,\text{cr}}$.
From the condition $E_3=E_{3,\text{cr}}$ and Eq.(\ref{eq_E3}), we find a quadratic equation of $\alpha_3$.
Solving its equation, we obtain
\beann
\alpha_3&=&0~\rm{or}~\alpha_3(s_1,s_2),
\\ 
\alpha_3(s_1,s_2)
	&:=&
	\frac{(2+s_2)(1-s_2-3 s_1 s_2+s_1^2 (2+s_2) )}{(1+2 s_1)(1-2s_2-2s_2^2)+s_1^2(2+s_2)}
\,.
\enann
Thus, we obtain the largest value of $E_3$ or $E_{3,\text{cr}}$ by inserting the solutions.
For $\alpha_3=0$, $E_3$ is equal to $E_1$. Hence, we find the maximal value of $E_3/(E_1+1)$;
\beann
\frac{E_3}{E_1+1}=\frac{1}{1+1/E_1} \le 0.4550 \quad (s_1=s^{\text{ISCO}}_{\text{min}})
\,.
\enann
On the other hand, for $\alpha_3=\alpha_3(s_1,s_2)$, 
we show the contour map of $E_3/(E_1+1)$ in Fig. \ref{Appendix_eff2_Elastic_isco}.
The maximum efficiency of 
$E_3/(E_1+1) \approx 0.06587$ is obtained at the red point $(s_1,s_2) \approx (0.08230, 0.449)$

\begin{figure}[h]
	\centering\includegraphics[width=7cm]{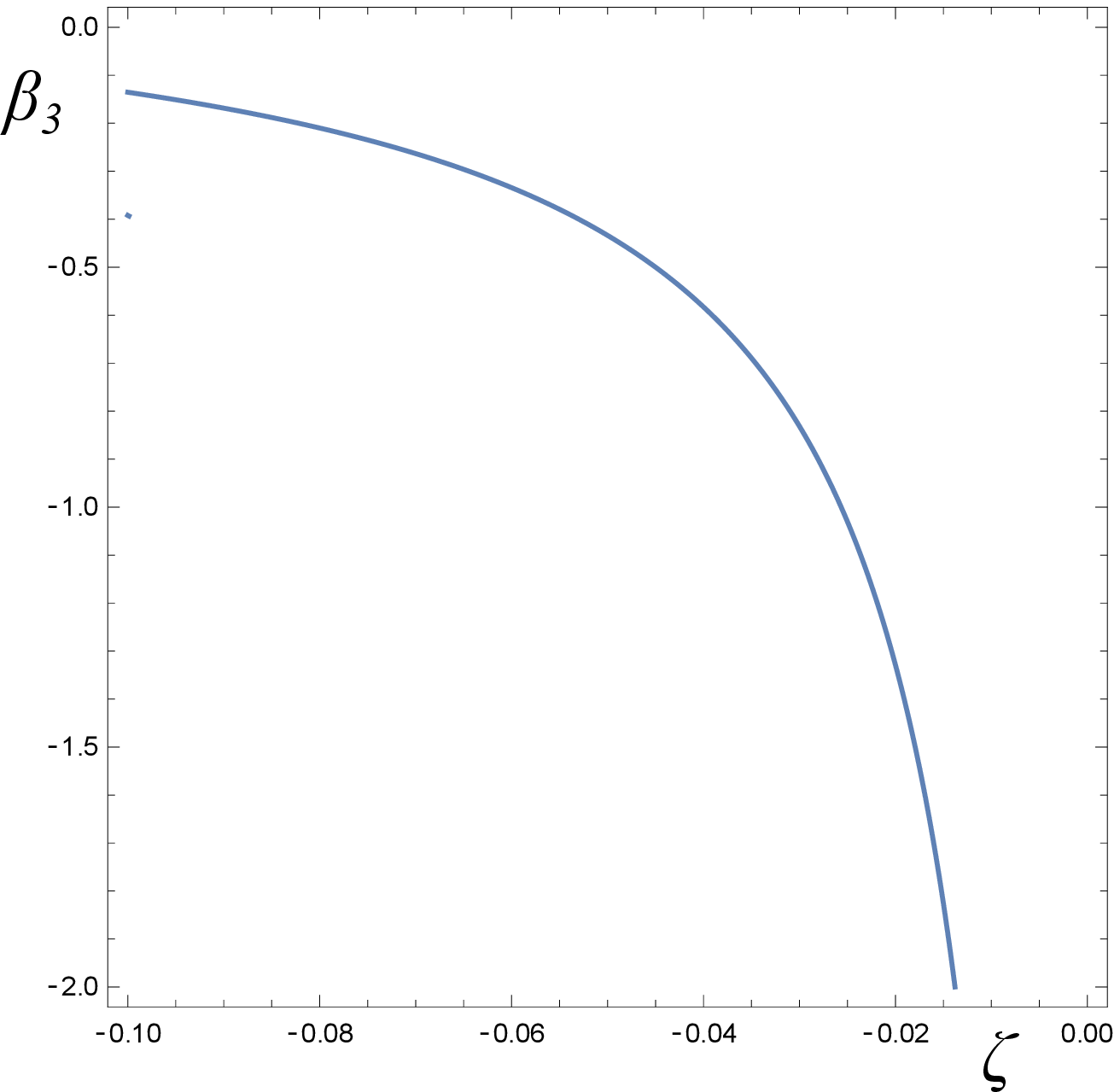}
\caption{The contour of $E_2=1$ in terms of $\zeta$ and $\beta_3$.
Here, $s_1=s^{\text{ISCO}}_{\text{min}}$ and we set $s_2=0$ as example.
In this case, the timelike condition for particle 2 is trivial and 
$\zeta$ can take an arbitrary value.
}
\label{Appendix_E2_Elastic_isco}
\end{figure}

Comparing these results, we find $E_3/(E_1+1) \approx 0.4550 \quad (s_1=s^{\text{ISCO}}_{\text{min}})$ 
as the maximum value for arbitrary $s_2$.
Since we find that $E_2=1$ is possible from Fig. \ref{Appendix_E2_Elastic_isco},
$\eta_{\rm max} \approx 0.4550$ is obtained as the maximum efficiency.

\subsection{Case[B] (MPM: photoemission)}
\begin{figure}[h]
	\centering
	\includegraphics[width=8cm]{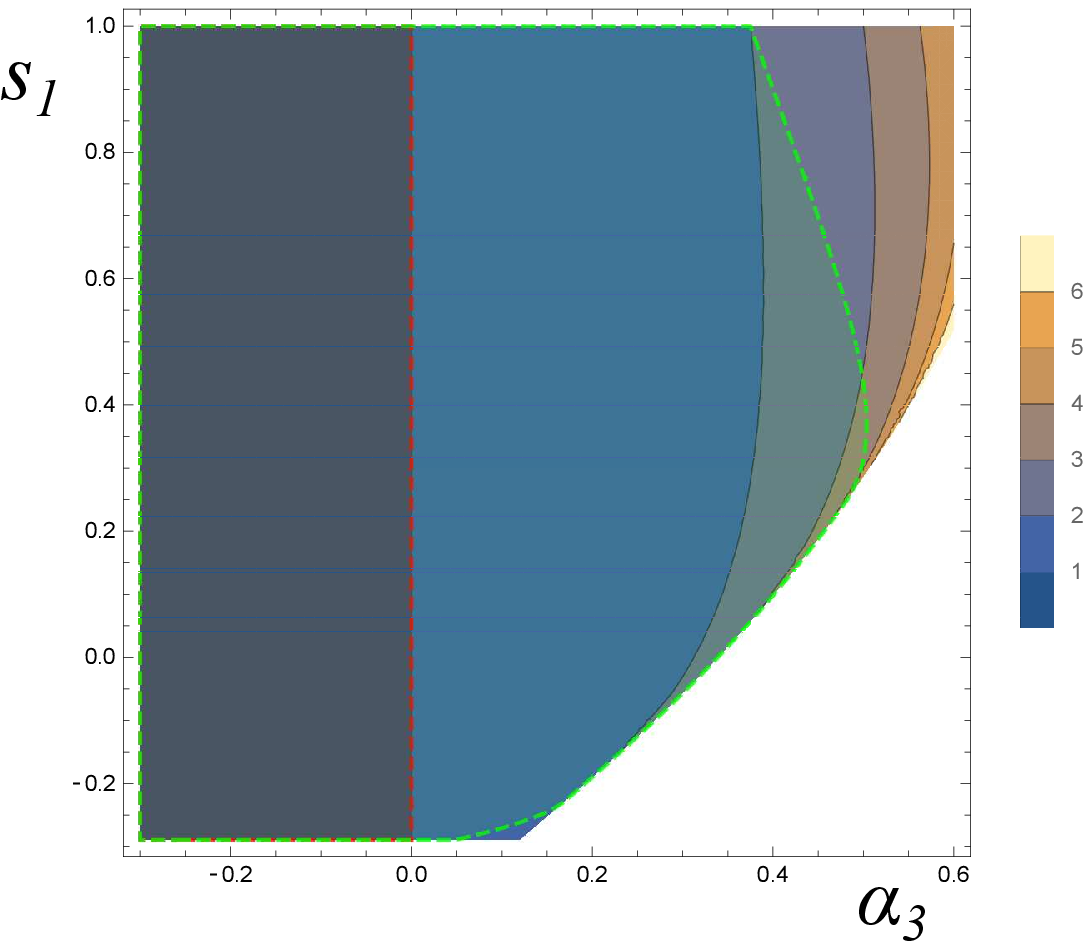}
\caption{The contour map of $E_3/E_1$ in terms of $\alpha_3$ and $s_1$.
In the light-green region, the timelike condition for the particle 3 orbit is satisfied 
In the light-red region, the condition $E_3 \le E_{3,({\rm cr})}$ is satisfied.
The maximum efficiency, $E_3/E_1 \approx 1.000$ is obtained at $\alpha_3 \approx 0$ for any $s_1$.
}
\label{Appendix_eff_Inverse_Compton_isco}
\end{figure}

In this case, the condition $E_3 \le E_{3,({\rm cr})}$ must be satisfied.
We show the contour map of $E_3/E_1$ in terms of $\alpha_3$ and $s_1$ in Fig. \ref{Appendix_eff_Inverse_Compton_isco}.
As the maximum efficiency, $E_3/E_1 \approx 1.000$ is obtained at $\alpha_3 \approx 0$ for any $s_1$.

This result has physical meaning when $E_2 = 0$ is possible in terms of $\zeta$ and $\beta_3$. 
To see this, from Eq. (\ref{calP}), we find that the asymptotic behavior of $\mathcal{P}$ becomes
\beann
{\cal P}\approx 
-8E_3 \zeta \beta_3\left[{E_3(2+s_1)\over (1-s_1)f(s_1, E_3, 0)}+1\right]
\,,
\enann
when we take the limit of  $\zeta \beta_3 \rightarrow -\infty$.
$\zeta$ is constrained as $-\infty<\zeta<0$ because particle 2 is non-spinning, while $\beta_3$ is arbitrary.
As a result, we obtain $E_2\rightarrow 0$ in the limit $\zeta \beta_3\rightarrow -\infty$. 
Hence, we find the maximum efficiency 
$\eta_{\rm max}\approx 1.000$ for the inverse Compton scattering.
For $s_1=0$, the maximum efficiency also becomes $\eta_{\text{max}}\approx 1.000$
since it does not depend on $s_1$.
\\
\bibliographystyle{IMANUM-BIB}
\bibliography{Collision_ISCO_spinning_ref_0330}

\end{document}